\documentclass[conference]{IEEEtran}
\IEEEoverridecommandlockouts
\usepackage{cite}
\usepackage{amsmath,amssymb,amsfonts}
\usepackage{algorithmic}
\usepackage{balance}
\usepackage{graphicx}
\usepackage{textcomp}
\usepackage{tabularx}
\usepackage{xcolor}
\usepackage{multirow}
\usepackage{subfigure}
\usepackage{url}
\def\BibTeX{{\rm B\kern-.05em{\sc i\kern-.025em b}\kern-.08em
    T\kern-.1667em\lower.7ex\hbox{E}\kern-.125emX}}

\title{BatSort: Enhanced Battery Classification with Transfer Learning for Battery Sorting and Recycling \\
\thanks{This work was supported in part by A*STAR under its MTC Programmatic (Award M23L9b0052), MTC Individual Research Grants (IRG) (Award M23M6c0113), the Ministry of Education, Singapore, under the Academic Research Tier 1 Grant (Grant ID: GMS 693), SIT’s Ignition Grant (STEM) (Grant ID: IG (S) 2/2023 – 792), and Future Communications Research \& Development Programme (FCP) under Grant FCP-SIT-TG-2022-007.}
}

\makeatletter
\newcommand{\linebreakand}{%
  \end{@IEEEauthorhalign}
  \hfill\mbox{}\par
  \mbox{}\hfill\begin{@IEEEauthorhalign}
}
\makeatother

\author{\IEEEauthorblockN{Yunyi Zhao, Wei Zhang$^*$\thanks{$^*$ Corresponding author. Email: wei.zhang@singaporetech.edu.sg.}}
\IEEEauthorblockA{
\textit{Singapore Institute of Technology}
\\\{yunyi.zhao, wei.zhang\}@singaporetech.edu.sg
}
\and
\IEEEauthorblockN{Erhai Hu, Qingyu Yan}
\IEEEauthorblockA{
\textit{Nanyang Technological University, Singapore}
\\\{huer0002, alexyan\}@ntu.edu.sg
}
\linebreakand 
\IEEEauthorblockN{Cheng Xiang}
\IEEEauthorblockA{
\textit{National University of Singapore}
\\elexc@nus.edu.sg
}
\and
\IEEEauthorblockN{King Jet Tseng}
\IEEEauthorblockA{
\textit{Singapore Institute of Technology}
\\kingjet.tseng@singaporetech.edu.sg
}
\and
\IEEEauthorblockN{Dusit Niyato}
\IEEEauthorblockA{
\textit{Nanyang Technological University, Singapore}
\\dniyato@ntu.edu.sg
}
}

\begin{document}
\bstctlcite{IEEEexample:BSTcontrol}

\maketitle

\begin{abstract}
Battery recycling is a critical process for minimizing environmental harm and resource waste for used batteries. However, it is challenging, largely because sorting batteries is costly and hardly automated to group batteries based on battery types. In this paper, we introduce a machine learning-based approach for battery-type classification and address the daunting problem of data scarcity for the application. We propose BatSort which applies transfer learning to utilize the existing knowledge optimized with large-scale datasets and customizes ResNet to be specialized for classifying battery types. We collected our in-house battery-type dataset of small-scale to guide the knowledge transfer as a case study and evaluate the system performance. We conducted an experimental study and the results show that BatSort can achieve outstanding accuracy of 92.1\% on average and up to 96.2\% and the performance is stable for battery-type classification. Our solution helps realize fast and automated battery sorting with minimized cost and can be transferred to related industry applications with insufficient data. 
\end{abstract}


\section{Introduction}
Batteries are portable carriers of electricity and play a crucial role in a wide spectrum of applications, encompassing domains from entertainment to transport and healthcare. However, their end-of-life management poses challenges. Many batteries contain environmentally harmful chemicals (e.g., mercury), which can potentially cause water and soil contamination \cite{gottesfeld2018soil}. Additionally, batteries consist of valuable chemical components (e.g., zinc, copper, and lithium) that are worth to be recycled \cite{miao2022overview}. Batteries therefore shall not be disposed of indiscriminately, emphasizing the significance of battery recycling for both sustainability and economic considerations \cite{yang2021sustainability}. Unfortunately, battery recycling is not straightforward, with only a small percentage of batteries being recycled globally. North America, for instance, disposes of 3.3 billion batteries annually without proper recycling \cite{website2}.

Battery sorting is a pivotal component of the recycling process. In this stage, batteries are collected and transported to a centralized facility where they are classified into different groups based on type. These sorted groups are then assigned to specific recycle bins, ensuring no mixing of batteries types. Such sorting process greatly promotes the efficiency of battery recycling, enabling a specialized and optimized recycling process for each battery type with batch processing. However, the common practice of such a sorting process predominantly relies on human-based rather than automated classification.

While there are existing studies on sorting, they are inapplicable to batteries. For instance, several machine learning (ML) algorithms have been evaluated for garbage sorting, e.g., convolutional neural network and region proposal networks \cite{zhihong2017vision}. However, the datasets for garbage are relatively large with thousands of images or more, and the classification problem is often simplified to consider main types like plastic and solid wastes with obvious shape differences. Conversely, battery sorting presents unique challenges. Take in-house batteries in our case study as an example: even for different brands of batteries, varying battery types can exhibit similar shapes, logos, and designs. Indeed, accurate battery sorting is not trivial. Several recent works have discussed battery recycling but from different perspectives, e.g., chemical process \cite{liu2021selective}\cite{kim2021comprehensive}. Battery sorting, especially automatic battery-type classification, has not been well studied. A high level of human intervention is still common and accurate ML models are still missing for battery-type classification, often attributed to the data scarcity.

In this paper, we propose a transfer learning-based solution for image-based battery-type classification for \underline{bat}tery \underline{sort}ing, named BatSort. To address the data scarcity issue, we leverage existing classification models from diverse applications, assuming that these models possess pertinent knowledge transferable to battery classification. By effectively employing this transferred knowledge, the reliance on data for achieving good classification performance shall be lower than solutions purely driven by battery data. Specifically, we choose a popular existing classification model as the backbone, which inherits the optimal parameters trained from large-scale datasets with millions of images. We re-configure the backbone by removing several old layers used for other applications and adding new layers to align with the configuration of our battery-type classification using our in-house battery type dataset. 
In summary, we make the following contributions in this paper:
\begin{itemize}
    \item Introduction of a system architecture for automatic battery sorting and the development of BatSort, which employs transfer learning for accurate battery-type classification with limited data;
    \item Compilation of an in-house dataset with over 500 images for 9 home batteries types, which is available to the research community \cite{batsort2024};
    \item Conducting an experimental study, which demonstrated that BatSort can achieve competitive battery-type classification performance with an average accuracy of 92.1\% and 2.03x improvement from a non-knowledge solution.
\end{itemize}

The benefits of the proposed BatSort is multi-sided. The direct benefit is the automation of battery-type classification which contributes to improving the system efficiency and reducing cost. Furthermore, BatSort serves as a beacon for other sectors, such as consumer electronics, that stand to gain from ML but are hamstrung by data limitations.

The rest of the paper is organized as follows. We present our system architecture of automated battery sorting in Section \ref{sec:sys-arch}. We describe our methodology of BatSort for battery-type classification in Section \ref{sec:method}. In Section \ref{sec:exp}, we present the experimental study and results for a case study of home batteries. Finally, we conclude this paper in Section \ref{sec:conclusion}.

\section{System Architecture and Description}
\label{sec:sys-arch}
\begin{figure}
    \centering	\includegraphics[width=0.98\linewidth]{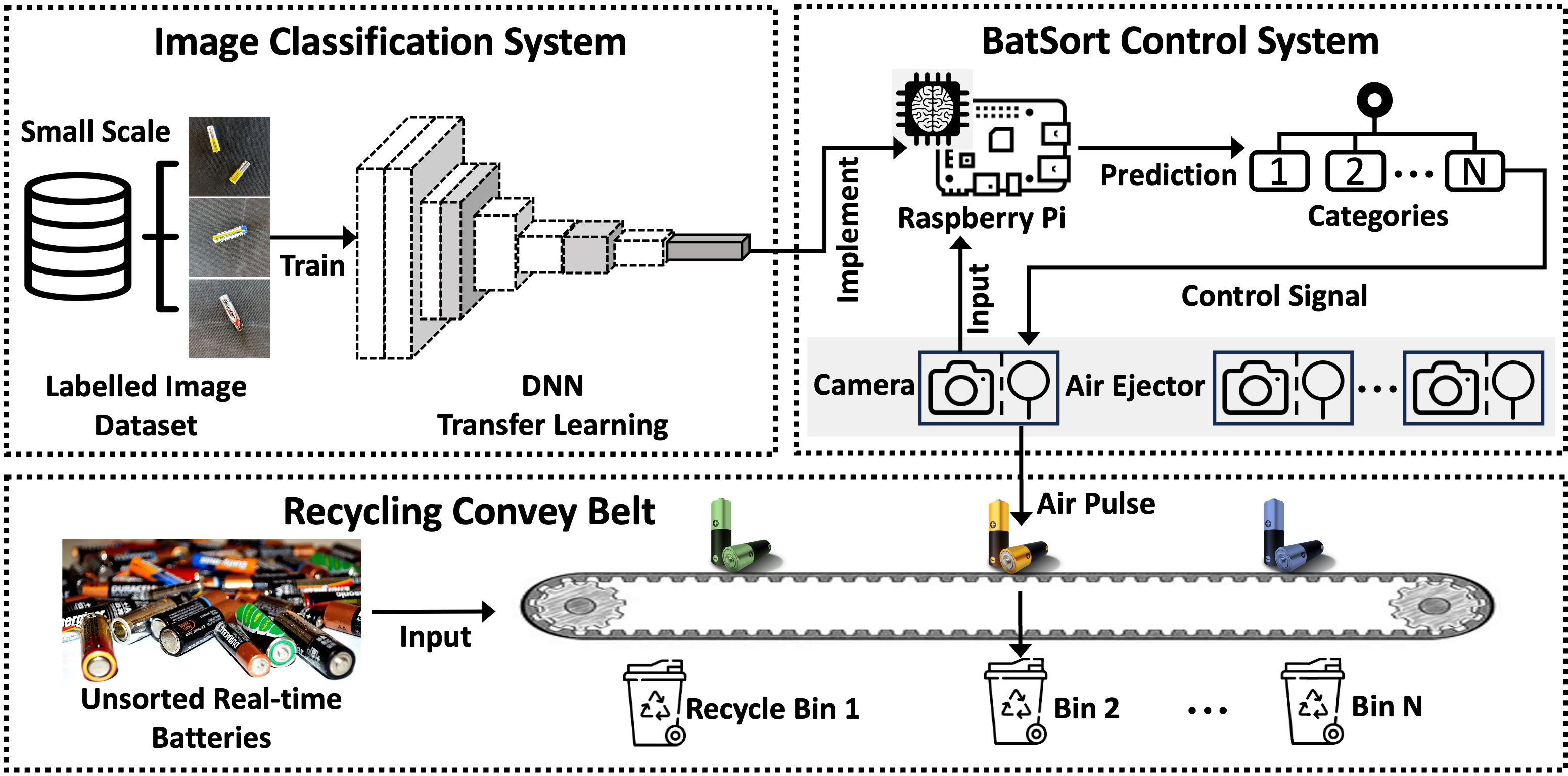}
    \caption{An illustration of the system architecture of battery sorting. Battery type can be determined automatically with BatSort and the batteries of the same type are grouped into the same recycling bins for further processing.}
    \label{fig:systemarch}
\end{figure}

We present the system architecture of automated battery sorting in this section, and its illustration is shown in Fig. \ref{fig:systemarch}. In a typical scenario, the system starts from a stream of batteries to be sorted and recycled. A conveyor belt carries those batteries and moves at a moderate speed. The goal is to pick out batteries of the same type (e.g., brand and chemicals) and place them into the same bins for further recycling processes, such as shredding and chemical processing. This goal, unfortunately, has been achieved mainly manually with huge labor costs and the competitiveness of our solution lies in the automatic and accurate battery-type classification and battery grouping. More details of our system are as follows.

\subsection{Battery Inspection and Conveyor Belt} 
Battery sorting in reality faces several challenges and the whole process includes several tasks. Batteries to be recycled are collected from different locations and stored in a centralized facility. Those collected batteries are often mixed with foreign objects like packages and wires, and they undergo a preliminary inspection to remove the foreign objects. Subsequently, the batteries are placed on a spacing conveyor to introduce a consistent distance between the batteries to facilitate the sorting process. Finally, the batteries are routed through a sorting conveyor and the sorting can be either manually or automated. In this research, we aim to expand existing hardware capability \cite{charpentier2023urban} with convey belt, controllers, etc., with our software capability, and achieve an automated system for battery sorting with minimized manpower overhead.

\subsection{Image-based Battery-type Classification}
The automated sorting system integrates both hardware and software components, featuring a camera and multiple air ejectors, each corresponding to a designated recycle bin for a specific battery type. The camera takes a photo of each battery on the conveyor belt and aligns its photo-taking frequency with the belt's moving speed. Each photo will be utilized in our image-based solution for battery-type classification. Upon identifying the battery type, the corresponding air ejector will trigger an air pulse to direct the battery into the its appropriate bin, to complete the sorting process of this specific battery. A key challenge here is in the software layer to detect the battery type correctly. The initial step involves image preprocessing, transforming the raw photos from the camera into a format suitable for classification model. The model takes in the image and predict the battery type, represented as a discrete label. 

Such an integrated solution with hardware and software can be realized in mainstream and cost-efficient edge devices such as Raspberry Pi. A database can be configured to store and manage the battery images and related information for further performance improvement, e.g., classification model fine-tuning. Besides, our system is designed to accommodate special circumstances. These include scenarios where a battery type is either new to our database, infrequently encountered, or when the model's prediction lacks sufficient confidence. For such instances, we propose to set up an additional recycling bin labeled \texttt{others} at the end of the conveyor. Batteries that do not trigger any ejector—owing to uncertain classification or unrecognized types—will naturally gravitate towards this bin. This mechanism ensures that all batteries, regardless of type or rarity, are systematically sorted and accounted for in our recycling process. For major changes in battery distribution (e.g., new brand and design), the software and bin deployment can also be updated without much additional workload.

\section{Methodologies for BatSort}
\label{sec:method}
Battery-type classification is the focus of this paper and the key to battery sorting. In this section, we propose our methodology for accurate battery-type classification using transfer learning. Same as many ML-based solutions, two building blocks are data and model, and we present them as follows.

\subsection{Data Collection and Pre-processing}
Data is one of the prerequisites for training an ML model. Nowadays, many image-based datasets are available online, e.g., ImageNet \cite{ILSVRC15} with 14 million images, but none of them, to our best knowledge, is specialized for battery-type classification. While batteries are included in some datasets, battery type information is not included. In this study, we conducted data collection and prepared an in-house battery-type database \cite{batsort2024}. We searched from various sources such as Google Images and online shopping websites with battery images and the image selection is based on different practical settings. Each image is manually labelled as 9 specific battery types and the label is verified by a different researcher before the image is included in the dataset. It is impractical to enumerate all battery types due to technical and cost constraints. In BatSort, we incorporate a comprehensive range of common battery types. For those outside this scope, we introduce a general category termed as \texttt{others}. Note that each image to be analyzed in the model is assigned a battery type with a quantifiable certainty, i.e., probability. Given the certainty is below a pre-defined threshold, e.g., 80\%, the input battery image will be labelled as \texttt{others}. 

The dataset is balanced with about 50 images for each battery type. We consider common battery types in Singapore, including Duracell (alkaline), IKEA (alkaline), Energizer (alkaline), Energizer (industrial), Energizer (lithium), Exell (Ni-MH), Exell (Ni-CD), GP (alkaline), and Klarus. We also have a group for miscellaneous \texttt{others} battery types with 50 images and they are used in model performance evaluation. Altogether we have $\sim$500 images. All images undergo data cleaning and pre-processing. This process entailed editing out irrelevant elements such as watermarks or source URLs that could potentially introduce biases in the ML model. The dataset is randomly divided into two parts for training and testing, with 80\% and 20\% images, respectively. In the training dataset, we allocate 10\% images for validation during training. Our dataset \cite{batsort2024} is available to the research community to advance the battery sorting and recycling research. In preparation for model training, images were uniformly resized to 244$\times$244$\times$3 with 244 for the image size and 3 for RGB channels. Armed with this dataset, we proceed to train our model.

\subsection{Transfer Learning-based Battery-Type Classification}
One of the daunting problems of many ML-based industry applications is data scarcity \cite{liu2022transline}. Compared to the mainstream ML datasets like ImageNet with millions of images, industrial datasets often are of much smaller scale, e.g., thousands or even hundreds of images. The scale is not (or far from) sufficient to train a well-performed ML model, which often involves millions of parameters to be optimized by learning from the data. Lightweight ML models with fewer parameters have been studied to relieve ML's data demand. However, these models usually cannot match the performance of their larger counterparts in most applications. Our battery-type classification is one of the applications facing the challenge of data scarcity. To mitigate this, we employ transfer learning, adapting the knowledge acquired from pre-trained, high-performing models in other applications to our specific task. Intuitively, such knowledge complements data and reduces a model's reliance on large volumes of data for effective battery classification. Indeed there exist well-performed general-purpose classification models to be utilized for our research. We introduce the main components below.

\subsubsection{Backbone Model}
The existing models to be utilized for knowledge transfer are referred to as backbone models. A backbone has a strong impact on the outcome of classification. A suitable backbone shall be capable enough for classification and meanwhile share sufficient similarity to the target application, e.g., battery-type classification. We evaluated several leading backbone models, ultimately selecting ResNet, trained on the ImageNet dataset, for its precise and consistent classification performance.  ResNet typically consists of convolutional layers, residual blocks, batch normalization, global average pooing, and fully connected output layers. A key feature of ResNet is to employ residual blocks with shortcut connections to address the vanishing gradient problem. We specifically adopted ResNet-50V2 \cite{he2016deep} with 50 neural network layers. This backbone has been deployed in many image classification applications with competitive performances.

\subsubsection{Transferred Knowledge} 
Given a trained backbone, the knowledge resides within its parameters, which are optimized by the training dataset. Often, the optimal parameters for one application are sub-optimal for another application, e.g., ImageNet-trained backbone cannot be applied to battery-type classification directly for achieving optimal performance. Despite years of effort in transfer learning, effective transfer of knowledge from its original application to a new application is still non-trivial and requires much customization. 

\paragraph{Backbone Reconfiguration}
The backbone trained from ImageNet is not specifically designed for battery classification. In a typical setting, the last layers of the backbone map the processed and abstracted information of an input image from the beginning layers to each considered class with a probability. Then, the most likely class, or the class with the highest probability, is directed to the output, indicating the label of the input image. For battery-type classification, we removed these ImageNet-specific layers and introduced new layers that correspond to the available battery types. In short, we preserve the backbone's beginning, also the majority, layers close to the input and customize the last layers close to the output for battery types. 

Specifically, let $\mathcal{D}_{s} = \{(x_{i}^s,y^s_{i})\}_{i=1}^n$ be the source domain ImageNet dataset with $n$ images and we label target domain dataset for battery as $\mathcal{D}_{t} = \{(x_{i}^t,y_{i}^t)\}_{i=1}^m$ with $m$ images. Naturally, the available labels in both domains do not fully overlap. We use the existing ResNet model $M_s(x^s;\theta_s^*)=y^s$ with optimal parameters $\theta_{s}^*$ and the model has three consecutive stages $M_s^\text{f}(\cdot;\theta_{s,\text{f}}^*)$, $M_s^\text{v}(\cdot;\theta_{s,\text{v}}^*)$, and $M_s^\text{a}(\cdot;\theta_{s,\text{a}}^*)$. We replace the last stage $M_s^\text{a}$ with application-specific $M_t^\text{a}$, producing the target model $M_t(x^t;\theta_{t,\text{f}},\theta_{t,\text{v}},\theta_{t,\text{a}})=y^t$, initialized as,
\begin{equation}
 \big(M_s^\text{f}(\cdot;\theta_{s,\text{f}}^*), M_s^\text{v}(\cdot;\theta_{s,\text{v}}^*), M_t^\text{a}(\cdot;\theta_{t,\text{a}}^0)\big),
\end{equation}
where the target model's first two stages are derived from the source model and the last stage is newly customized. The model also inherits the the source model's optimal parameters $\theta_{s,\text{f}}^*$ and $\theta_{s,\text{v}}^*$ as $\theta_{t,\text{f}}$ and $\theta_{t,\text{v}}$, respectively, for the first two stages and initializes the parameters of $M_t^\text{a}$ randomly as $\theta_{s,\text{a}}^0$.

\paragraph{Parameter Optimization} 
The reconfigured model for battery-type classification contains the parameters optimized for ImageNet in the beginning layers and new, randomized parameters in the final layers. We can maintain or fine-tune the beginning layers' parameters and search for the optimal parameters in the last layers. Commonly in transfer learning, the beginning layers are considered to be responsible for detecting low-level features and patterns, e.g., corners, edges, and shapes. Intermediate layers also play an important role in deep feature extraction and the layers that are closer to the output increasingly become specialized and significant to the application. For ResNet, the shortcut connections play an essential role in preserving the low-level features and enhancing the gradient flow \cite{he2016deep}. Those shortcut connections help the reconfigured backbone adapt to domain shifts effectively where the low-level features are minimally affected by domain differences. Given the above observations and assumptions, we fixed the beginning layers' parameters to be non-trainable and only train the final layers that are close to the output. We expect BatSort's newly configured layers to map generic image features extracted from the beginning layers to the battery-type classification settings, e.g., battery colors, shapes, and patterns.

Specifically for the target model $M_t(\cdot;\theta_{t,\text{f}},\theta_{t,\text{v}},\theta_{t,\text{a}})$, we fix the parameters $\theta_{t,\text{f}}$ as $\theta_{s,\text{f}}^*$ for the first stage and make the rest two stages trainable/changeable. Essentially, we search for the optimal $\theta_{t,\text{v}}^*$ and $\theta_{t,\text{a}}^*$ for the last two stages as,
\begin{equation}
(\theta_{t,\text{v}}^*,\theta_{t,\text{a}}^*) = \arg\min_{\theta_{t,\text{v}},\theta_{t,\text{a}}} \mathcal{L}(D_t; \theta_{t,\text{v}},\theta_{t,\text{a}}),
\end{equation}
where $\mathcal{L}(\cdot)$ is the cross-entropy loss function. Finally, the optimal target model can be represented as $M_t(\cdot;\theta_{t}^*)$ where the parameters consist of $\theta_{s,\text{f}}^*$, $\theta_{t,\text{v}}^*$, and $\theta_{t,\text{a}}^*$, respectively, for the three stages, with optimal battery classification performance.

\section{Experimental Study}
\label{sec:exp}
We evaluate the effectiveness of BatSort for battery-type classification in this section. We will present the details of data collection and categorization, backbone settings, and experimental results as follows.


\subsection{Backbone Configuration and Experimental Setup}
\begin{figure}
    \centering	\includegraphics[width=0.9\linewidth]{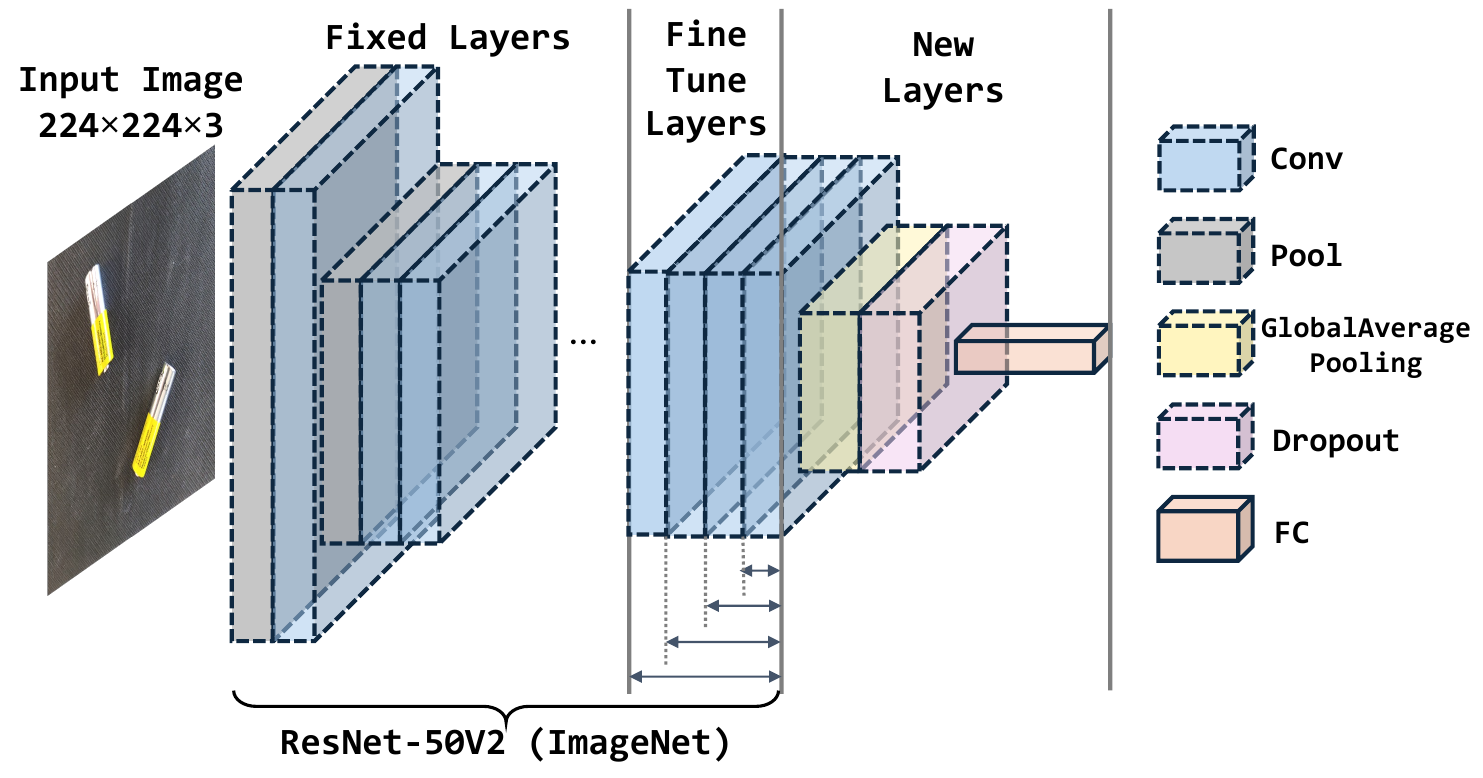}
    \caption{An illustration of our transfer learning-based BatSort model with the last layers of ResNet-50V2 replaced with three new layers for customized classification and only the last seven layers of the new model trainable.
    }
    \label{fig:network}
\end{figure}

We present BatSort's backbone configuration and settings first. We use ResNet-50V2 with parameters optimized with ImageNet. The new customized layers include a \texttt{GlobalAveragePooling2D} layer for pooling, a \texttt{Dropout} layer with a dropout rate of 20\%, and a \texttt{Dense} layer with 9 neurons for the 9 battery types in our dataset. We adopt a 2-stage training strategy with a large learning rate of 0.005 for fast search in the first stage and a slow rate of 0.00005 for fine-grained optimization in the second stage. The first stage has 300 training epochs and the second stage has 100 epochs with early-stopping activated when the validation accuracy does not improve for 10 epochs. In the first stage, we focus on optimizing the weights of the 3 new layers as well as a few adjacent layers, as shown in Fig. \ref{fig:network}. In the second stage, we assume the parameters for the last layers are near-optimal and we fine-tune two more adjacent layers with inherited parameters. All experiments are run in a workstation with an AMD Ryzen 9 5950X processor and NVIDIA GTX 3080 GPU and the reported experimental results are based on 10 independent runs for cross-validation. For each independent run of the experiment, we re-generate the training and testing dataset randomly. Worth mentioning that our methodology is versatile and extendable in dealing with other battery categories, e.g., from cars or electronic consumables, which however are beyond this paper's scope.

\subsection{Knowledge Transfer Performance}
We evaluate the knowledge transfer effectiveness for BatSort in this part. The key idea of BatSort is to inherit ResNet and its optimized parameters as knowledge. Such knowledge is expected to be important to complement the data for battery sorting, and we investigate the knowledge's quantitative impact here. Two natural benchmarks for such investigation are the backbone with randomized parameters and the backbone with ImageNet-optimized parameters without customization. The former is a common practice for ML-based applications with collecting data and training models from the beginning, i.e., with randomized initial parameters. The latter is a brute-force knowledge utilization schema assuming that the knowledge can be applied for target applications without or with minimal customization, e.g., optimizing the parameters only for newly added layers. We present the accuracy of battery-type classification for the benchmarks and BatSort in Table \ref{tab:knowledge}.

\begin{table}
\centering
\caption{The knowledge transfer effectiveness of BatSort in terms of battery classification accuracy and improvement with respect to mean accuracy compared to two benchmarks where one does not use knowledge and the other uses the knowledge in a non-optimal way.}
\renewcommand{\arraystretch}{1.3}
\begin{tabular}{ccccccc}
\hline\hline
\multirow{2}{*}{Model}       & \multicolumn{4}{c}{Accuracy}\\ \cline{2-5} 
         & Mean     & Best     & SD  & Improv.   \\ \hline
No knowledge &   30.4\%   &  51.0\%  & 0.114 & 0\%      \\ \hline
Non-optimal knowledge &  83.1\%  &  88.7\%  &  0.048  & 1.73x \\ \hline
BatSort  &  \textbf{92.1\%} &   \textbf{94.3\%}   &  \textbf{0.024}   &   \textbf{2.03x}   \\
\hline\hline
\end{tabular}
\label{tab:knowledge}
\end{table}
\begin{table}
\renewcommand{\arraystretch}{1.3}
\centering
\caption{BatSort's performance sensitivity to the number of trainable layers beside the 3 newly added layers in the backbone in terms of battery-type classification accuracy. The gap to the optimal mean accuracy is reported for each number which shall be neither too large nor too small.}
\begin{tabular}{ccccc}
\hline\hline
\multirow{2}{*}{No. of Trainable Layers}         & \multicolumn{4}{c}{Accuracy} \\ \cline{2-5} 
         & Mean     & Best     & SD   & Gap  \\ \hline
0 &   83.1\%   &  88.7\%  & 0.048 & 9.0\% \\ \hline
1 &   90.2\%   &  94.2\%  & 0.042 & 1.9\%\\ \hline
2  &  \textbf{92.1\%} &   \textbf{94.3\%}   & \textbf{0.024} & $-$ \\ \hline
3 &   84.0\%   &  92.5\%  & 0.051 & 8.1\% \\ \hline
4 &   69.0\%   &  86.8\%  & 0.114 & 23.1\%\\
\hline\hline
\end{tabular}
\label{tab:trainable-layers}
\end{table}

\subsubsection{Comparison to No Knowledge}
Table \ref{tab:knowledge} shows that BatSort outperforms the two benchmarks significantly. It achieves 92.1\% classification accuracy on average and the accuracy can be up to 94.3\%. When all the parameters are randomized initially and no knowledge is utilized, the accuracy drops to 30.4\% on average and even the best run only achieves 51.0\% accuracy. With respect to the average performance, BatSort is 1.73x better than the first benchmark. The implication is that knowledge is important and shall be utilized to achieve optimal performance for battery-type classification.

\subsubsection{Comparison to Non-Optimal Knowledge}
The second benchmark utilizes existing knowledge for battery-type classification. Despite the knowledge being optimized for a different application, the results show that it is still valuable for our application. On average, the accuracy is improved by 1.73x compared to the no knowledge benchmark and the top accuracy can be close to 90\%. However, the knowledge is not utilized optimally in this benchmark, where the beginning layers of the backbone maintain the same parameters without fine-tuning and customization and only the last newly added layers and parameters are optimized with our collected dataset. With more effective knowledge transfer, BatSort is 10.8\% more accurate than the benchmark. This highlights the importance of customizing existing knowledge and optimizing the parameters beyond the new layers for battery-type classification.

\subsubsection{Model Stability}
We also report the results of standard deviation (SD). Small SD implies stable model performance with minimal accuracy variation in different independent runs. We can observe that BatSort is not only more accurate but also more stable than the benchmarks. Specifically, BatSort's SD is 0.024, which is only 21\% and 50\% of the two benchmark's SDs, respectively. Knowledge does contribute to improved performance stability beyond accuracy and even the non-optimal parameters help reduce the SD by over a half.

\subsubsection{Application Complication and Summary}
battery-type classification itself is a challenging application with many battery types. Given a list of the types, BatSort ranks the probability of all types (e.g., the summation is 100\%) for a given battery image. Only the highest probability is chosen to label the battery type of the image. Intuitively, being the highest among more classes is more difficult. For example, in a power line anomaly detection application with two classes \cite{liu2022transline}, the accuracy can reach near 80\% even if no knowledge is used and improves to over 96\% with knowledge transfer. The complication highlights the importance of transfer learning in our battery-type classification with many classes.

In summary, the experimental results on knowledge transfer demonstrate that BatSort's usage of transfer learning is effective for battery-type classification. Such usage is not trivial and shall be well-designed, and the benefits of successful knowledge transfer are improved accuracy and stability.

\subsection{Sensitivity Analysis}
The optimal knowledge transfer relies on the optimization of the model and training. We specifically investigate the performance sensitivity to trainable layers and the dropout rate.

\subsubsection{Trainable Layers}
We have discussed different ways of knowledge utilization above and demonstrated that knowledge is useful for battery-type classification and knowledge customization is necessary. In this part, we investigate the optimal knowledge transfer in fine granularity. We follow our training settings presented in Section \ref{sec:method}. The beginning layers of the backbone inherit the existing parameters which are unchanged during training, i.e., those layers are not trainable. The final newly added 3 layers are trainable and the parameters are optimized during the training from randomized parameters initially. The problem here is how many layers between the beginning layers and the last layers shall be trainable, assuming that those layers are more application-specific than the beginning layers of the backbone. We vary the number of such trainable layers from 0 to 4 to observe its impact on the classification accuracy of BatSort. For all the tests in this part, the dropout rate is fixed as 20\%.

We present the results in Table \ref{tab:trainable-layers}. The best performance is achieved when two layers before the new layers are trainable, where the classification accuracy is 92.1\% on average and up to 94.3\%, same as the results in Table \ref{tab:knowledge}. With fewer layers trainable, the accuracy drops. Reducing the number from 2 to 1, the accuracy is about 2\% lower than the optimal setting. Further reducing the number to 0, meaning none layers are trainable beside the new layers, the accuracy is even lower with a 9\% gap to the optimal. Increasing the number beyond 2 shows a similar pattern. The performance decreases with an 8\% gap from the optimal by adding one additional trainable layer and decreases further, e.g., 23\% gap, with more added. Overall, the knowledge cannot be well customized to the target application if too many layers and the knowledge is non-customized, and the data scale of our battery sorting application is not large enough to well train many layers.

\subsubsection{Dropout Rate}
One common problem for data scarcity in ML is over-fitting. This is especially true when the ML model is large and the data scale cannot match the scale and capacity of the model. As a result, the models can be too specialized for the training dataset and perform sub-optimally for non-training data after training. We introduce a dropout layer \cite{kong2022reflash} in our customized backbone for regularization. The layer randomly deactivates a certain percentage of the neurons during the training to prevent the neurons from being over-fitted collectively and the percentage is referred to as dropout rate. We expect the dropout layer to help our backbone to be generalized to the battery images not in the training dataset and we aim to find the optimal dropout rate for our application.

\begin{figure}
     \centering
     \def \tmpw{0.45}
    \subfigure[Training]{\includegraphics[width=\tmpw\linewidth]{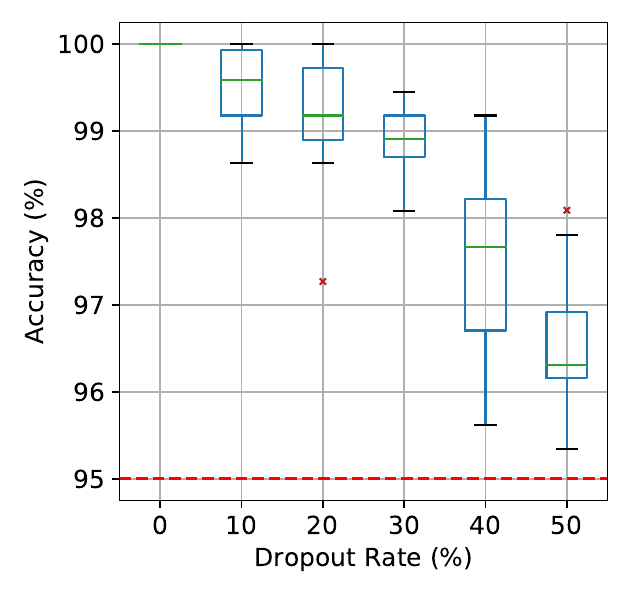}\label{fig:dropout:training}}
    \subfigure[Testing]{\includegraphics[width=\tmpw\linewidth]{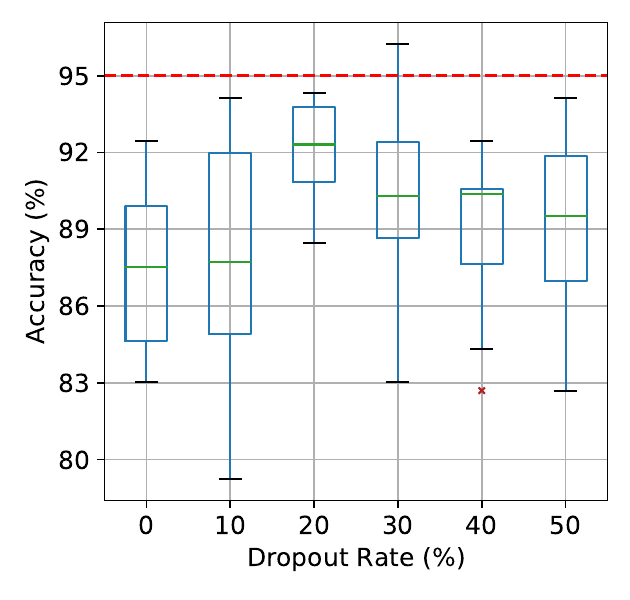}\label{fig:dropout:testing}}
    \caption{The BatSort's performance sensitivity in terms of accuracy to dropout rate, which varies between 0\% to 50\%, for both training and testing stages of the battery-type classification model. The red dashed line marks 95\% accuracy for easy comparison. A high box means good accuracy and a small box implies stable performance. The optimal dropout rate with the best average accuracy and stability for battery-type classification is 20\%.}
    \label{fig:dropout}
\end{figure}

We test a list of dropout rates from 0\% to 50\% and report the performance in Fig. \ref{fig:dropout}. With a higher rate, more neurons in the dropout layer are deactivated and it becomes more challenging for the model to be specialized for the training data. Fig. \ref{fig:dropout:training} and Fig. \ref{fig:dropout:testing} show the performance of the classification model training and testing stages, respectively. Comparing the results in the two figures, the over-fitting is relatively evident, with the accuracy in the testing stage generally lower than in the training stage. Over-fitting is especially true for a small dropout rate. For example, when the rate is 10\%, the accuracy is very close to 100\% for training data, and the performance downgrade is significant for testing data with below 90\% accuracy. Increasing the rate moderately makes the classification model less over-fitted to the training data and more generalized to the testing data. The highest accuracy in our experiments is achieved with a 30\% dropout rate and the top accuracy is 96.2\%. But the overall performance with a 30\% dropout rate is not the most competitive with big accuracy variation in different runs. The optimal rate in our experiments is 20\%, where the average accuracy for testing data is the highest at 92.1\% with a small SD of 0.024. For the rest tested rates such as 50\%, the model's performance is outperformed by the optimal setting for both training and testing data and the performance degradation is more significant with larger rates. Overall, the dropout layer is useful to alleviate the over-fitting problem for classification with data scarcity, and the optimal rate shall be neither too small nor large.

\section{Conclusion}
\label{sec:conclusion}
\balance 
In this paper, we introduced BatSort, a transfer learning-based solution, for automatic battery-type classification, a key step in efficient battery sorting and recycling. Recognizing the lack of dedicated ML research in this area, we curated a specialized dataset for in-house batteries with over 500 images for 9 battery types for our case study. By leveraging ResNet's parameters, we demonstrated how integrating a modestly sized dataset with strategic knowledge transfer substantially enhances the accuracy of battery classification. Based on our experimental study, we achieved a classification accuracy of 92.1\% on average and up to 96.2\%, significantly outperforming the non-transfer learning approaches by 2.03x and showing a more than 10\% improvement over non-optimized knowledge transfer. Overall, BatSort not only brings a high degree of precision and reliability to the automation of battery sorting but also has potential applications in various other battery categories, offering insights for similar industrial scenarios.


\section*{Acknowledgement}
The authors acknowledge the contribution of the students from Singapore Institute of Technology for preliminary study in their applied learning modules CSC2008 and CSC3001.

\bibliographystyle{IEEEtran}
\bibliography{reference}

\begin{thebibliography}{10}
\providecommand{\url}[1]{#1}
\csname url@samestyle\endcsname
\providecommand{\newblock}{\relax}
\providecommand{\bibinfo}[2]{#2}
\providecommand{\BIBentrySTDinterwordspacing}{\spaceskip=0pt\relax}
\providecommand{\BIBentryALTinterwordstretchfactor}{4}
\providecommand{\BIBentryALTinterwordspacing}{\spaceskip=\fontdimen2\font plus
\BIBentryALTinterwordstretchfactor\fontdimen3\font minus
  \fontdimen4\font\relax}
\providecommand{\BIBforeignlanguage}[2]{{%
\expandafter\ifx\csname l@#1\endcsname\relax
\typeout{** WARNING: IEEEtran.bst: No hyphenation pattern has been}%
\typeout{** loaded for the language `#1'. Using the pattern for}%
\typeout{** the default language instead.}%
\else
\language=\csname l@#1\endcsname
\fi
#2}}
\providecommand{\BIBdecl}{\relax}
\BIBdecl

\bibitem{gottesfeld2018soil}
P.~Gottesfeld, F.~H. Were, L.~Adogame \emph{et~al.}, ``Soil contamination from
  lead battery manufacturing and recycling in seven african countries,''
  \emph{Environmental research}, vol. 161, pp. 609--614, 2018.

\bibitem{miao2022overview}
Y.~Miao, L.~Liu, Y.~Zhang \emph{et~al.}, ``An overview of global power
  lithium-ion batteries and associated critical metal recycling,''
  \emph{Journal of Hazardous Materials}, vol. 425, p. 127900, 2022.

\bibitem{yang2021sustainability}
Y.~Yang, E.~G. Okonkwo, G.~Huang \emph{et~al.}, ``On the sustainability of
  lithium ion battery industry--a review and perspective,'' \emph{Energy
  Storage Materials}, vol.~36, pp. 186--212, 2021.

\bibitem{website2}
\BIBentryALTinterwordspacing
Recycle right: What you need to know about end-of-life batteries. [Online].
  Available: \url{https://www.emterra.ca/blogs/green-factor}
\BIBentrySTDinterwordspacing

\bibitem{zhihong2017vision}
C.~Zhihong, Z.~Hebin, W.~Yanbo \emph{et~al.}, ``A vision-based robotic grasping
  system using deep learning for garbage sorting,'' in \emph{2017 36th Chinese
  control conference (CCC)}.\hskip 1em plus 0.5em minus 0.4em\relax IEEE, 2017,
  pp. 11\,223--11\,226.

\bibitem{liu2021selective}
F.~Liu, C.~Peng, Q.~Ma \emph{et~al.}, ``Selective lithium recovery and
  integrated preparation of high-purity lithium hydroxide products from spent
  lithium-ion batteries,'' \emph{Separation and Purification Technology}, vol.
  259, p. 118181, 2021.

\bibitem{kim2021comprehensive}
S.~Kim, J.~Bang, J.~Yoo \emph{et~al.}, ``A comprehensive review on the
  pretreatment process in lithium-ion battery recycling,'' \emph{Journal of
  Cleaner Production}, vol. 294, p. 126329, 2021.

\bibitem{batsort2024}
\text{[Online]},
  \url{https://github.com/FriedrichZhao/Singapore_Battery_Dataset}.

\bibitem{charpentier2023urban}
N.~M. Charpentier, A.~A. Maurice, D.~Xia \emph{et~al.}, ``Urban mining of
  unexploited spent critical metals from e-waste made possible using advanced
  sorting,'' \emph{Resources, Conservation and Recycling}, vol. 196, p. 107033,
  2023.

\bibitem{ILSVRC15}
O.~Russakovsky, J.~Deng, H.~Su \emph{et~al.}, ``{ImageNet Large Scale Visual
  Recognition Challenge},'' \emph{International Journal of Computer Vision
  (IJCV)}, vol. 115, no.~3, pp. 211--252, 2015.

\bibitem{liu2022transline}
F.~Liu, T.~W. Low, W.~Zhang \emph{et~al.}, ``Transline: Transfer learning for
  accurate power line anomaly detection with insufficient data,'' in \emph{ICC
  2022-IEEE International Conference on Communications}.\hskip 1em plus 0.5em
  minus 0.4em\relax IEEE, 2022, pp. 5543--5548.

\bibitem{he2016deep}
K.~He, X.~Zhang, S.~Ren \emph{et~al.}, ``Deep residual learning for image
  recognition,'' in \emph{Proceedings of the IEEE conference on computer vision
  and pattern recognition}, 2016, pp. 770--778.

\bibitem{kong2022reflash}
X.~Kong, X.~Liu, J.~Gu \emph{et~al.}, ``Reflash dropout in image
  super-resolution,'' in \emph{Proceedings of the IEEE/CVF Conference on
  Computer Vision and Pattern Recognition}, 2022, pp. 6002--6012.

\end{thebibliography}

\end{document}